\documentclass[reqno]{amsart}

\usepackage{booktabs} 
\usepackage{IEEEtrantools}
\usepackage{amssymb,latexsym,amsfonts,amsmath}
\usepackage{graphicx}
\usepackage{mathrsfs}
\usepackage{dsfont}

\usepackage{subfigure}

\usepackage{tikz}
\usetikzlibrary{calc,shapes,arrows}
\usepackage{subfig}

\usepackage{algorithm}
\usepackage{algorithmic}

\usepackage{xspace}

\newtheorem{theorem}{Theorem}[section]
\newtheorem{lemma}[theorem]{Lemma}

\newtheorem{problem}[theorem]{Problem}

\newtheorem{definition}[theorem]{Definition}

\newtheorem{remark}[theorem]{Remark}
\newtheorem{assumption}{Assumption}
\numberwithin{equation}{section}

\newcommand{\R}{{\mathbb{R}}}

\newcommand{\N}{{\mathbb{N}}}

\newcommand{\PP}{\mathds{P}}

\usepackage[many]{tcolorbox}
\usetikzlibrary{calc}
\tcbuselibrary{skins}

\newtcolorbox{resp}[1][]{%
	enhanced jigsaw,%
	colback=gray!5!white,%
	colframe=gray!80!black,%
	size=small,%
	boxrule=1pt,%
	halign title=flush center,%
	coltitle=black,%
	breakable,%
	drop shadow=black!50!white,%
	attach boxed title to top left={xshift=1cm,yshift=-\tcboxedtitleheight/2,yshifttext=-\tcboxedtitleheight/2},%
	minipage boxed title=3cm,%
	boxed title style={%
		colback=white,%
		size=fbox,%
		boxrule=1pt,%
		boxsep=2pt,%
		underlay={%
			\coordinate (dotA) at ($(interior.west) + (-0.5pt,0)$);
			\coordinate (dotB) at ($(interior.east) + (0.5pt,0)$);
			\begin{scope}[gray!80!black]
				\fill (dotA) circle (2pt);
				\fill (dotB) circle (2pt);
			\end{scope}
		}%
	},%
	#1%
}

\usepackage{fancyhdr}

\newenvironment{nouppercase}{%
	\renewcommand{\uppercasenonmath}[1]{}}{}

\begin{document}

\begin{abstract}
In this work, we propose a data-driven approach for the construction of finite abstractions (\emph{a.k.a.,} symbolic models) for discrete-time deterministic control systems with unknown dynamics. We leverage notions of so-called \emph{alternating bisimulation functions} (ABF), as
a relation between each unknown system and its symbolic model, to quantify the mismatch between state behaviors of two systems. Accordingly, one can employ our proposed results to perform formal verification and synthesis over symbolic models and then carry the results back over unknown original systems. In our data-driven setting, we first cast the required conditions for constructing ABF as a robust optimization program (ROP). Solving the provided ROP is not tractable due to the existence of unknown models in the constraints of ROP. To tackle this difficulty, we collect finite numbers of data from trajectories of unknown systems and propose a scenario optimization program (SOP) corresponding to the original ROP. By establishing a probabilistic relation between optimal values of SOP and ROP, we formally construct ABF between unknown systems and their symbolic models based on the number of data and a required confidence level. We verify the effectiveness of our data-driven results over two physical case studies with unknown models including (i) a DC motor and (ii) a \emph{nonlinear} jet engine compressor. We construct symbolic models from data as appropriate substitutes of original systems and synthesize policies maintaining states of unknown systems in a safe set within infinite time horizons with some guaranteed confidence levels.
\end{abstract}

\title{{\Large Data-Driven Synthesis of Symbolic Abstractions with Guaranteed Confidence}$^*$\footnote[1]{$^*$This work was supported in part by the Swiss National Science Foundation under NCCR Automation, grant agreement 51NF40-180545.}}

\author{{\bf {\large Abolfazl Lavaei}}}
\author{{\bf {\large Emilio Frazzoli}}\\
{\normalfont Institute for Dynamic Systems and Control, ETH Zurich, Switzerland}\\
\texttt{\{alavaei,efrazzoli\}@ethz.ch}}

\pagestyle{fancy}
\lhead{}
\rhead{}
\fancyhead[OL]{Abolfazl Lavaei and Emilio Frazzoli}

\fancyhead[EL]{Data-Driven Synthesis of Symbolic Abstractions with Guaranteed Confidence} 
\rhead{\thepage}
\cfoot{}

\begin{nouppercase}
	\maketitle
\end{nouppercase}

\section{Introduction}

{\bf Motivations.} Formal methods have received remarkable attentions, over the past two decades, as a promising approach to analyze complex dynamical systems while providing mathematical guarantees. Examples of such complex systems span a broad range of real-life safety-critical
applications including (air) traffic networks, robotic
manufacturing, biological systems, automated vehicles, etc.,
to name a few. Providing formal verification and controller synthesis framework for this type of complex systems is inherently very challenging mainly due to their computational complexity arising from uncountable sets of states and inputs. 

To mitigate the encountered computational difficulties, one potential solution, proposed in the relevant literature, is to construct finite abstractions (\emph{a.k.a.,} symbolic models) as approximate descriptions of continuous-space systems in which each finite state correlates with a group of continuous states of original (concrete) systems. 
One can then leverage the constructed finite abstractions as a suitable substitution of original systems, perform verification or controller synthesis over abstract models, and ultimately carry the results back over original complex systems. Since the mismatch between original systems and their finite abstractions exists within a guaranteed error bound, one can ensure that original systems also fulfill the same property of interest as finite abstractions.
Two different types of symbolic abstractions have been proposed in the relevant literature: (i) \emph{sound} abstractions whose behavior embraces that of concrete systems and (ii) \emph{complete} abstractions whose behavior is identical to that of original systems~\cite{tabuada2009verification}. In particular, constructing complete abstractions provides a \emph{sufficient and necessary guarantee}: 
there exists a controller satisfying a desired property of interest over symbolic abstractions \emph{if and only if} there exists a controller fulfilling the same property in the concrete domain. On the downside, existence of a sound abstraction only results in a \emph{sufficient guarantee}: failing to synthesize a controller for a property of interest over the sound abstraction does not prevent the existence of a controller for the concrete system.

{\bf State of the Art.} There has been a comprehensive literature on the abstraction-based analysis of deterministic control systems. Construction of (in)finite abstractions, for different classes of deterministic models,
has been broadly studied in~\cite{tabuada2009verification,girard2009approximately,le2013mode,girard2015safety,coogan2015efficient}, to name a few.  In order to deal with the \emph{curse of dimensionality} problem as the main challenge in the construction of finite abstractions, \emph{compositional} abstraction-based techniques have been proposed in the past few years in~\cite{tazaki2008bisimilar,pola2016symbolic,zamani2017compositional,mallik2018compositional,swikir2019compositional}, to name a few. 

Although the existing results on the construction of finite abstractions are promising, unfortunately, all of them require knowing precise models of the systems; and hence,
they are not applicable when the underlying model is unknown.
In particular, since closed-form models for many physical systems are either not available or too complex to be of any practical use, one cannot employ model-based techniques to analyze those complex dynamical systems.
Although there exist some results based on \emph{indirect data-driven} techniques to provide analysis frameworks for unknown dynamical systems by learning approximate models via identification approaches (\emph{e.g.,}~\cite[and references herein]{Hou2013model}), acquiring an accurate model is always very challenging, time-consuming, and expensive, especially if the underlying dynamics are too complex which is the case in many real-world applications. This critical challenge motivated us to bypass the system identification phase and develop a \emph{direct data-driven} approach to construct finite abstractions directly by collecting data from trajectories of unknown original systems.

{\bf Original Contributions.}
Our main contribution in this work is to propose a data-driven approach for the construction of symbolic models for unknown deterministic control systems while providing formal guarantees. We utilize
a notion of \emph{alternating bisimulation functions} (ABF) to relate each unknown system to its finite abstraction and quantify the mismatch between state behaviors of two systems. 
To do so, we reformulate the required conditions for constructing ABF as a
	robust optimization program (ROP). We then collect finite numbers of data from trajectories of unknown systems and propose a scenario optimization program (SOP) associated to its original ROP. We eventually establish a probabilistic relation between optimal values of SOP and ROP, and formally construct ABF between unknown systems and their symbolic models with a required confidence level. 
We apply our data-driven approaches to two physical case studies with unknown models including (i) a DC motor and (ii) a \emph{nonlinear} jet engine compressor.
We provide proofs of all statements in Appendix.

{\bf Relevant Works.} A learning-based approach based on Gaussian process regression for the construction of symbolic models is presented in~\cite{hashimoto2020learning}. 
The proposed approach assumes that the underlying system consists of both known and unknown dynamics and the main goal is to learn unknown models via Gaussian processes.
In comparison with our setting, we propose here a \emph{direct data-driven} approach to construct finite abstractions from unknown models \emph{without} performing any system identifications. A probably approximately correct (PAC) statistical framework for the data-driven construction of finite abstractions is proposed in~\cite{devonport2021symbolic}. The proposed results in~\cite{devonport2021symbolic} are only applicable for the construction of \emph{sound} abstractions (\emph{sufficient guarantee}), whereas our data-driven approaches construct \emph{complete} abstractions (\emph{sufficient and necessary guarantee}). 
Accordingly, we provide an alternating \emph{bisimulation} relation between original and abstract models whereas~\cite{devonport2021symbolic} only establishes a one-side alternating simulation relation from a symbolic model to its original system.
Moreover, the proposed results in~\cite{devonport2021symbolic} only handle finite-time horizon properties since their symbolic model is constructed with a confidence level and a violation threshold that increases with the time horizon. In comparison, our data-driven symbolic model is built with confidence $1$ and can be utilized for infinite-time horizon properties, as well. Data-driven safety verification of stochastic systems is recently proposed in~\cite{Ali_ADHS21} but using \emph{barrier certificates} (if existing) rather than constructing symbolic abstractions (always existing) which is the case in our work.

\section{Discrete-Time Deterministic Control Systems}\label{Sec: dt-NDS}

\subsection{Notation and Preliminaries}

Sets of real, positive and non-negative real numbers are denoted by $\mathbb{R},\mathbb{R}^+$, and $\mathbb{R}^+_0$, respectively. We denote the sets of non-negative and positive integers by $\mathbb{N} := \{0,1,2,...\}$ and $\mathbb{N}^+=\{1,2,...\}$, respectively. Given $N$ vectors $x_i \in \mathbb{R}^{n_i}$, $x=[x_1;...;x_N]$ denotes the corresponding column vector of dimension $\sum_i n_i$. Minimum and maximum eigenvalues of a symmetric matrix $A$ are denoted by $\lambda_{\min}(A)$ and $\lambda_{\max}(A)$, respectively. Given sets $X$ and $Y$, a relation $\mathscr{R}\subseteq X \times Y$ is a subset of the Cartesian product $X \times Y$ that relates $x \in X$ to $y \in Y$ if $(x, y) \in \mathscr{R}$, equivalently denoted by $x\mathscr{R}y$. The absolute value of $a\in\mathbb R$ is denoted by $\vert a\vert$. We denote the Euclidean norm of a vector $x\in\mathbb{R}^{n}$ by $\Vert x\Vert$. For any matrix $P\in\mathbb R^{m\times n}$, we have $\|P\| := \sqrt{\lambda_{\max}(P^\top P)}$. A function $\varphi: \mathbb{R}^+_0 \rightarrow \mathbb{R}^+_0$ is said to be a class $\mathcal{K}$ function if it is continuous, strictly increasing, and $\varphi(0)=0$. A class $\mathcal{K}$ function $\varphi(\cdot)$ belongs to class $\mathcal{K}_\infty$ if $\varphi(s) \rightarrow \infty$ as $s \rightarrow \infty$. Given a probability space $(\mathcal D,\mathbb B(\mathcal D),\PP)$, we denote by $\mathcal D^N$ the $N$-Cartesian product of set $\mathcal D$, and by $\PP^N$ its corresponding product measure. The operator $\vDash$ is employed to show the feasibility of a solution for an optimization problem.

\subsection{Discrete-Time Deterministic Control Systems}
We consider discrete-time deterministic control systems (dt-DCS) in this work as formalized in the following definition. 

\begin{definition}
	A discrete-time deterministic control system (dt-DCS) is represented by the tuple
	\begin{align}\label{EQ:1}
	\Upsilon=(X,U,f),
	\end{align}
	where:
	\begin{itemize}
		\item $X\subseteq \mathbb R^n$ is the state set of the system;
		\item $U = \{\nu_1,\nu_2,\dots,\nu_m\}$ with $\nu_i\in\mathbb R^{p}, i \in\{1,\dots,m\}$ is the discrete input set of the system;
		\item $f:X\times U\rightarrow X$ is a measurable function
		characterizing the state evolution of the system, which is assumed to be \emph{unknown}.
	\end{itemize}
\end{definition}
Evolution of the state of dt-DCS $\Upsilon$ for a given initial state $x(0)\in X$ and an input sequence $\nu(\cdot):\mathbb N\rightarrow U$ is described as
\begin{align}\label{EQ:2}
\Upsilon\!:x(k+1)=f(x(k),\nu(k)),\quad k\in\mathbb N. 
\end{align} 
We employ $x_{a\nu}:\mathbb N\rightarrow X$ to denote the \emph{state trajectory} of $\Upsilon$ at time $k\in\mathbb N$ under an input sequence $\nu(\cdot)$ starting from an initial condition $x(0)= a$.

\subsection{Symbolic Models}\label{Symbolic}

In this work, we approximate a dt-DCS $\Upsilon$ with a \emph{finite} abstraction, \emph{a.k.a.,} symbolic model, with a discrete state set~\cite{pola2016symbolic}. 
To construct such a finite approximation, we assume that the state space of the dt-DCS $\Upsilon$ is restricted to a compact subset over which we are interested to perform analysis. The abstraction algorithm first constructs finite partitions of the state set as $X = \cup_i \mathsf X_i$
and then selects representative points $\bar x_i\in \mathsf X_i$ as abstract states~\cite{pola2016symbolic,swikir2019compositional}.

Given a dt-DCS $\Upsilon=(X,U,f)$, the symbolic model $\hat\Upsilon$ can be constructed as
\begin{equation*}
\hat\Upsilon =(\hat X, U, \hat f),
\end{equation*}
where $\hat X := \{\bar x_i, i=1,...,n_{\bar x}\}$ is the finite state set of~$\hat\Upsilon$. Moreover, $\hat f:\hat X \times U\rightarrow\hat X$ is defined as 
\begin{equation}\label{EQ:3}
\hat f(\hat{x}, \nu) = \mathcal Q(f(\hat{x}, \nu)),
\end{equation}
where the map $\mathcal Q\!:X\rightarrow \hat X$ allocates to any $x\in X$, a representative point $\bar x\in\hat X$ of the corresponding partition set and satisfies the inequality
\begin{equation}\label{EQ:4}
\Vert \mathcal Q(x)-x\Vert \leq \delta,\quad \forall x\in X, 
\end{equation}
with $\delta:=\sup\{\|x-x'\|,\,\, x,x'\in \mathsf X_i,\,i=1,2,\ldots,n_{\bar x}\}$ being the \emph{state discretization parameter}.

In the next section, we present a notion of \emph{alternating bisimulation functions}, as a relation between dt-DCS $\Upsilon$ and its symbolic model $\hat\Upsilon$, to quantify the mismatch between state behaviors of two systems.

\section{Alternating Bisimulation Functions}\label{Sec:ABF}
Here, we present alternating bisimulation functions (ABF) as formalized in the following definition~\cite{swikir2019compositional}.

\begin{definition}\label{Def:1}
	Consider a dt-DCS $\Upsilon =(X,U,f)$ and its 
	symbolic model $\hat\Upsilon =(\hat X,U,\hat f)$. A function $\mathcal V:X\times\hat X\to\R_0^+$ is called an alternating bisimulation function (ABF) between $\hat\Upsilon$ and $\Upsilon$ if there exist
	$\sigma\in\R^+$, $0< \gamma <1,$ and $\rho \in\R_0^+$ such that
	for all $x\in X$, $\hat x\in\hat X$, $\nu\in U$, one has
	\begin{align}\label{EQ:5}
	\sigma\Vert x - \hat x\Vert^2&\leq \mathcal V(x,\hat x),\\\label{EQ:6}
	\mathcal V(f(x,\nu),\hat{f}(\hat x,\nu))&\leq\max\big\{\gamma\mathcal V(x,\hat{x}),\rho\big\}.
	\end{align}
	If there exists an ABF $\mathcal V$ between $\hat\Upsilon$ and $\Upsilon$, we denote it by $\hat\Upsilon\cong_{\mathcal{A}}\Upsilon$.
\end{definition}

\begin{remark}
	Note that we assumed $\sigma$ in~\eqref{EQ:5} is linear in $\Vert x - \hat x\Vert^2$ for the sake of a simpler presentation. However, one can readily extend it to be a $\mathcal{K}_{\infty}$ function but at the cost of introducing extra decision variables in ROP~\eqref{ROP}, and accordingly, solving SOP~\eqref{SOP} with more number of data (cf. Theorem~\ref{Thm:3}).
\end{remark}

Next theorem, borrowed from~\cite{swikir2019compositional}, shows the usefulness of ABF to quantify the mismatch between state behaviors of dt-DCS $\Upsilon$ and its symbolic model $\hat\Upsilon$.

\begin{theorem}\label{Thm:1}
	Consider a dt-DCS $\Upsilon =(X,U,f)$ and its 
	symbolic model $\hat\Upsilon =(\hat X,U,\hat f)$. Suppose $\mathcal V$ is an ABF between $\hat\Upsilon$ and $\Upsilon$ as in Definition~\ref{Def:1}, \emph{i.e.,} $\hat\Upsilon\cong_{\mathcal{A}}\!\Upsilon$. Then a relation $\mathscr{R} \subseteq X \times \hat X$ defined by
	\begin{align}\label{Relation}
	\mathscr{R} := \Big\{(x,\hat x) \in X \times \hat X \,\big|\, \mathcal V(x,\hat{x}) \leq\rho\Big\},
	\end{align}
	is an $\tilde\epsilon$-approximate alternating bisimulation relation~\cite{tabuada2009verification} between $\hat\Upsilon$ and $\Upsilon$ with 
	\begin{align*}
	\tilde\epsilon = (\frac{\rho}{\sigma})^{\frac{1}{2}}.
	\end{align*}
\end{theorem}\vspace{0.1cm}

We now state the main problem that we aim to solve in this work.

\begin{resp}
	\begin{problem}\label{Prob:1}
		Consider the dt-DCS $\Upsilon$ in~\eqref{EQ:2} with an unknown transition map $f$. Develop a data-driven approach for the construction of ABF, as a relation between unknown $\Upsilon$ and its symbolic model $\hat\Upsilon$, with a-priori confidence bound $\beta \in[0,1]$ as
		\begin{align*}
		\PP^{\mathcal N}\Big\{\hat\Upsilon\cong_{\mathcal{A}}\!\Upsilon\Big\}\ge 1-\beta.
		\end{align*}\vspace{-0.4cm}
	\end{problem}
\end{resp}
To address Problem~\ref{Prob:1}, we propose our data-driven framework in the next section.

\section{Data-Driven Construction of ABF}\label{DD-ABF}

In this section, we assume that the transition map $f$ in~\eqref{EQ:2} is unknown and the main goal is to construct an ABF using data. In our data-driven setting, we take two consecutive data-points from trajectories of unknown $\Upsilon$ as the pair of $(x(k),x(k+1))$ and denote it by $(x_i, f(x_i,\nu))$. We also fix the structure of ABF as the form $\mathcal{V}(\eta,x,\hat x)=\sum_{j=1}^{z} {\eta}_jq_j(x,\hat x)$ with some user-defined (possibly nonlinear) basis functions $q_j(x,\hat x)$ and unknown coefficients $\eta=[{\eta}_{1};\ldots;\eta_z] \in \mathbb{R}^z$. It is worth mentioning that our proposed techniques do not put any restrictions on the type of the basis functions in the structure of ABF. For instance, in the case of polynomial-type ABF, basis functions $q_j(x,\hat x)$ are monomials over $x,\hat x$.

In order to enforce required conditions for the construction of ABF as~\eqref{EQ:5}-\eqref{EQ:6}, we first cast the problem as the following robust optimization program (ROP):
\begin{align}\label{ROP}
&\text{ROP}\!:\left\{
\hspace{-1.5mm}\begin{array}{l}\min\limits_{[\Psi;\mu]} \quad\mu,\\
\, \text{s.t.} \quad  \,\max_j\big\{\phi_j(x, \hat x,\nu, \Psi)\big\}\leq \mu,  j\in\{1,2\}, \\ 
\quad\quad\quad\!\forall x\in X, ~\forall \hat x\in \hat X, ~\forall \nu\in U,\\
\quad\quad\quad\!\Psi = [\sigma;\tilde\gamma;\tilde\rho;{\eta}_{1};\dots;\eta_z],\\
\quad\quad\quad\!\sigma\in\R^+, \tilde\gamma \in (0,1), \tilde\rho\in\R_0^+, \mu\in\mathbb R,\end{array}\right.
\end{align}
where: 
\begin{align}\notag
\phi_1(x,\hat x,\nu, \Psi)& = \sigma\Vert x - \hat x\Vert^2 - \mathcal V(\eta,x,\hat x),\\\label{EQ:11}
\phi_2(x,\hat x,\nu, \Psi)& = \mathcal V(\eta,f(x,\nu),\hat{f}(\hat x,\nu))- \tilde\gamma\mathcal V(\eta,x,\hat{x})-\tilde\rho.
\end{align}
We denote the optimal value of ROP by $\mu_{\mathcal R}^*$. If $\mu_{\mathcal R}^* \leq 0$, a solution to the ROP implies the satisfaction of conditions~\eqref{EQ:5}-\eqref{EQ:6} for the construction of ABF.

\begin{remark}
	Note that condition $\phi_{2}$ is not convex due to a bilinearity between decision variables $\eta$ and unknown variable $\tilde\gamma$. To deal with this non-convexity, we assume that $\tilde\gamma$ lives in a finite set with a cardinality $l$, \emph{i.e.,} $\tilde\gamma \in \{\tilde\gamma_1,\dots,\tilde\gamma_l\}$. We then utilize the cardinality $l$ in computing the minimum number of data required for solving our optimization problem (cf. Theorem~\ref{Thm:3}). 
\end{remark}

\begin{remark}
	Since condition~\eqref{EQ:6} is presented in the $\max$-form, we reformulated it in $\phi_2$ as an implication-form. Then $\gamma, \rho$ in~\eqref{EQ:6} can be recovered based on $\tilde\gamma,\tilde\rho$ in $\phi_2$ as $\gamma = 1 - (1 - \psi)(1 - \tilde\gamma),\rho = \frac{\tilde \rho}{(1 - \tilde\gamma)\psi}$, for any $0<\psi <1$.
\end{remark}
To solve the proposed ROP in~\eqref{ROP}, one needs to know the precise map $f$ which is unknown in our setting. To tackle this problem, we propose our data-driven solution as the following. Let $(x_i)^{\mathcal N}_{i=1}$ denote $\mathcal N$ independent-and-identically distributed (i.i.d.) sampled data within $X$. Instead of solving the ROP in~\eqref{ROP}, we solve the following scenario optimization program (SOP):

\begin{align}\label{SOP}
&\text{SOP}\!:\left\{
\hspace{-1.5mm}\begin{array}{l}\min\limits_{[\Psi;\mu]} \,\,\,\,\,\mu,\\
\, \text{s.t.}  \quad \,\max_j\big\{\phi_j(x_i,\hat x,\nu,\Psi)\big\}\leq \mu, j\in\{1,2\}, \\
\quad \quad\quad \!\!\forall x_i\in X,\forall i\in \{1,\ldots,\mathcal N\}, \forall \hat x\in \hat X,\forall \nu\in U,\\
\quad\quad\quad\!\!\Psi = [\sigma; \tilde\gamma;\tilde\rho;{\eta}_{1};\dots;\eta_z],\\
\quad\quad\quad\!\sigma\in\R^+, \tilde\gamma \in (0,1), \tilde\rho\in\R_0^+, \mu\in\mathbb R,\end{array}\right.
\end{align}
where $\phi_1, \phi_2$ are the same functions as defined in~\eqref{EQ:11}. We denote the optimal value of SOP by $\mu_{\mathcal N}^*$. One can readily observe that~\eqref{SOP} has a finite number of constraints of the same form as in~\eqref{ROP}. Moreover, one can substitute $f(x_i,\nu)$ in $\phi_2$ by measurements of unknown dt-DCS after one-step evolution starting from $x_i$ under an input $\nu$.  

\begin{remark}
	Note that for computing $\hat f(\hat x,\nu)$ in $\phi_2$, one needs to first initialize the black-box model from $\hat x$ and feed an input $\nu$ to obtain $f(\hat x,\nu)$ as its one-step transition. Given a state discretization parameter $\delta$, $\hat f(\hat x,\nu)$ is then acquired as the \emph{nearest representative point} to the value of $f(\hat x,\nu)$ satisfying condition~\eqref{EQ:4}. This way of constructing symbolic abstractions from data introduces an approximation error that can be structured in $\rho$ in condition~\eqref{EQ:6}. We refer the interested reader to~\cite{swikir2019compositional} in which the closed-form approximation error $\rho$ is constructed based on $\delta$.
\end{remark}

\begin{remark}
	Since the collected data required for solving SOP~\eqref{SOP} should be i.i.d., one is allowed to take only one paired sample $(x_i, f(x_i,\nu))$ from each trajectory of unknown systems.
\end{remark}

In the next section, we establish a probabilistic relation between optimal values of SOP (\emph{i.e.,} $\mu_{\mathcal N}^*$) and ROP (\emph{i.e.,} $\mu_{\mathcal R}^*$).

\section{Data-Driven Guarantee for ABF}

Here, inspired by~\cite{esfahani2014performance}, we establish a formal relation between optimal values of $\text{SOP}$ and $\text{ROP}$. Accordingly, we formally construct an ABF $\mathcal V$ between symbolic model $\hat\Upsilon$ and unknown original system $\Upsilon$ based on the number of data and a required confidence level. To do so, we first raise the following assumption.

\begin{assumption}\label{Assm:1}
	Suppose $\phi_1,\phi_2$ are Lipschitz continuous with respect to $x$ with, respectively, Lipschitz constants $\mathcal I_{1}$, $\mathcal I_{{2_k}}$, for given $\tilde\gamma_k$ where $\ k\in\{1,\dots,l\}$.
\end{assumption} 

Under Assumption~\ref{Assm:1}, we propose the next theorem as the main result of this section.

\begin{theorem}\label{Thm:3}
	Consider an unknown dt-DCS in~\eqref{EQ:1}. Let Assumption~\ref{Assm:1} hold. Consider the $\text{SOP}$ in~\eqref{SOP} with its associated optimal value $\mu^*_{\mathcal N}$ and solution $\Psi^* = [\sigma^*;\tilde\rho^*,{\eta}^*_{1};\dots;\eta^*_z]$, with ${\mathcal N}\geq \mathcal {\bar N}\big(\bar\varepsilon,\beta\big)$, $\bar\varepsilon := (\bar\varepsilon_{1}, \dots, \bar\varepsilon_{l})$, where 	
	\begin{align}\label{EQ:12}
	\mathcal N(\bar\varepsilon,\beta)\!:=\min\Big\{\mathcal N\in\N \,\big|\,\sum_{k=1}^{l}\sum_{i=0}^{r-1}\binom{\mathcal N}{i}\bar\varepsilon^i_k(1-\bar\varepsilon_k)^{\mathcal N-i}\leq\beta\Big\},
	\end{align}
	$\beta\in [0,1]$ and $\bar\varepsilon_k=(\frac{\varepsilon_k}{\mathcal I_{\phi_k}})^{n}$, where $\varepsilon_k\in [0,1]\leq \mathcal I_{\phi_k} = \max\{\mathcal I_{1}, \mathcal I_{{2_k}}\}$, with $n,r,l$ being, respectively, dimension of the state set, number of decision variables in $\text{SOP}$~\eqref{SOP}, and cardinality of a finite set that $\tilde\gamma$ is taking value from it. If $$\mu^*_{\mathcal N} + \max_k\varepsilon_k \leq 0,$$ then the constructed $\mathcal V$ via solving $\text{SOP}$~\eqref{SOP} is an ABF between $\hat\Upsilon$ and $\Upsilon$ with a confidence of at least $1-\beta$, i.e.,
	\begin{align*}
	\PP^{\mathcal N}\big\{\hat\Upsilon\cong_{\mathcal{A}}\Upsilon\big\}\ge 1-\beta.
	\end{align*}
\end{theorem}

\begin{remark}
	Our proposed approach here provides a more tractable way of constructing ABF, compared to model-based techniques, by solving a mix-integer linear programing rather than, for instance, a mix-integer semi-definite programing using sum-of-squares (SOS) optimization problem (\emph{i.e.,} equivalently a mix-integer semi-infinite linear programming) but at the cost of providing a confidence bound over the ABF construction. Although constructing an ABF from model-based techniques is more accurate since no confidence is involved, one can push the proposed confidence to be close to 1 at the cost of collecting more data. In particular, if the number of data goes to infinity, the confidence converges to 1. In addition, our proposed data-driven technique is not only applicable to polynomial-type dynamics which is required in SOS optimization problems, but also to more complex unknown systems in which the ROP problem does not have any theoretical solution.
\end{remark}

\begin{remark}
	Note that the minimum number of data in~\eqref{EQ:12} required for solving SOP~\eqref{SOP} is exponential with respect to the dimension of unknown systems. However, the main benefit of our technique compared to system identification is that the proposed data-driven approach here is capable of constructing ABF for \emph{any type of nonlinear systems which are Lipschitz continuous}, whereas system identification approaches are mainly tailored to linear or some particular classes of nonlinear systems. In addition, even if one is able to find a model using system identification techniques, one still needs to search for ABF. In this case, one suffers from the computational complexity in both identifying the model as well as searching for ABF based on it.
\end{remark}

In Algorithm~\ref{Alg:1}, we summarize the required procedure for the data-driven construction of ABF.
	\begin{algorithm}[h]
		\caption{Data-driven construction of ABF}
		\label{Alg:1}		
		\begin{center}
			\begin{algorithmic}[1]
				\STATE Select a-priori $\varepsilon_k,\beta \in[0,1]$ as desired, with $\varepsilon_k\leq \mathcal I_{\phi_k}$ 
				\STATE Compute the minimum required number of data as $\mathcal N\geq \mathcal {\bar N}\big((\frac{\varepsilon_k}{\mathcal I_{\phi_k}})^{n},\beta\big)$
				\STATE Solve $\text{SOP}$~\eqref{SOP} with the acquired data and obtain $\mu^*_{\mathcal N}$
				\STATE If $\mu^*_{\mathcal N} + \max_k\varepsilon_k\leq0$, then the constructed $\mathcal V$ from solving $\text{SOP}$ is an ABF between $\hat\Upsilon$ and $\Upsilon$  with a confidence of at least $1-\beta$, \emph{i.e.,}
				$\PP^{\mathcal N}\big\{\hat\Upsilon\cong_{\mathcal{A}}\!\Upsilon\big\}\ge 1-\beta.$
			\end{algorithmic}
		\end{center}
\end{algorithm}

In order to compute the required number of data in Theorem~\ref{Thm:3}, one needs to first compute $\mathcal I_{\phi_k}$. In the following lemmas, we propose an explicit way to compute $\mathcal I_{\phi_k}$ for the choice of \emph{quadratic} ABF for both linear and nonlinear dt-DCS.

\begin{lemma}\label{Lem:1}
	Consider a linear dt-DCS $x(k+1)=Ax(k) + B\nu(k)$. For a quadratic ABF of the form $(x-\hat x)^\top P(x-\hat x)$, with a positive-definite matrix $P\in\mathbb{R}^{n\times n}$, $\mathcal I_{\phi_k}$ is computed as $\mathcal I_{\phi_k} \!= \max \big\{\mathcal I_{1}, \mathcal I_{{2_k}}\big\}$ with 
	\begin{align*}
	\mathcal I_{1} &= 4\alpha_1 (\lambda_{\min}(P) + \lambda_{\max}(P)),\\
	\mathcal I_{{2_k}} &= 2\lambda_{\max}(P) (2\mathscr I_1^2 \alpha_1 + 2\mathscr I_1 \mathscr I_2 \alpha_2 + \mathscr I_1\delta + 2\alpha_1\tilde\gamma_k),
	\end{align*}
	where $\Vert A\Vert \leq \mathscr I_1\in\mathbb R_0^+$, $\Vert B\Vert \leq \mathscr I_2\in\mathbb R_0^+$, $\Vert x\Vert \leq \alpha_1\in\mathbb R_0^+$ for any $x\in X$, $\Vert \nu\Vert \leq \alpha_2 \in\mathbb R_0^+$ for any $\nu\in U$.
\end{lemma}

\begin{remark}\label{Gerschgorin}
	Note that one needs to know upper bounds for $\lambda_{\max}(P)$ in order to compute $\mathcal I_{\phi_k}$ and the required number of data. The pre-assumed upper bounds should be then enforced as some additional conditions while solving the $\text{SOP}$ in~\eqref{SOP}. In particular, we enforce the required conditions as linear bounds on entries of the matrix $P$ using Gershgorin circle theorem \cite{varga2010gervsgorin} to provide an upper bound for $\lambda_{\max}(P)$, and accordingly, an over-approximation for $\mathcal I_{\phi_k}$ (cf. case studies).
\end{remark}	

We now propose another lemma for the computation of $\mathcal I_{\phi_k}$ for \emph{nonlinear} dt-DCS.

\begin{lemma}\label{Lem:1_1}
	Consider a nonlinear dt-DCS as in~\eqref{EQ:2}. Assume that $\Vert f(x,\nu)\Vert \leq \mathscr I_f\in\mathbb R_0^+$ and $\Vert \partial_{x}f(x,\nu) \Vert = \Vert \frac{\partial f(x,\nu)}{\partial x}\Vert \leq \mathscr I_{x}\in\mathbb R_0^+$. Then $\mathcal I_{\phi_k}$ for a quadratic ABF of the form $(x-\hat x)^\top P(x-\hat x)$, with a positive-definite matrix $P\in\mathbb{R}^{n\times n}$, is computed as $\mathcal I_{\phi_k} \!= \max \big\{\mathcal I_{1}, \mathcal I_{{2_k}}\big\} = \max\big\{2\lambda_{\max}(P) (2\mathscr I_f\mathscr I_x + \mathscr I_f\delta + 2\alpha_1\tilde\gamma_k),4\alpha_1 (\lambda_{\min}(P) + \lambda_{\max}(P))\big\}$, where $\Vert x\Vert \leq \alpha_1\in\mathbb R_0^+$ for any $x\in X$.
\end{lemma}

\begin{remark}
	Note that Lipschitz constant of dynamics and an upper bound on unknown dynamics are the minimal information we need from the system in our data-driven setting in Lemmas~\ref{Lem:1}, \ref{Lem:1_1}. However, one can utilize the proposed approach in~\cite{wood1996estimation} to estimate the Lipschitz constant of dynamics via sampled data. As for an upper bound on unknown dynamics, it can be readily computed based on the range of the state set. It is worth mentioning that some physical considerations about unknown models can be useful to find insights on the class of underlying systems which is required in Lemmas~\ref{Lem:1}, \ref{Lem:1_1}. As an alternative way, one can always leverage the proposed bound for the nonlinear models in Lemma~\ref{Lem:1_1} as it is also valid for linear systems.
\end{remark}

\section{Case Studies}\label{Sec:Case}

{\bf DC Motor.} We first apply our data-driven approaches to a DC motor adapted from~\cite{adewuyi2013dc} as follows:
\begin{align*}
x_1(k+1) &= x_1(k) + \tau\big (-\frac{R}{L}x_1(k) - \frac{k_{dc}}{L} x_2(k) + 0.7 \nu_1(k)\big ),\\
x_2(k+1) &= x_2(k) + \tau\big (\frac{k_{dc}}{J}x_1(k) - \frac{b}{J} x_2(k) + 0.7 \nu_2(k)\big ),
\end{align*}
where $x_1, x_2, R = 1, L = 0.01$, and $J = 0.01$ are, respectively, the armature current, the rotational speed of the shaft, the electric resistance, the electric inductance, and the moment of inertia of the rotor. In addition, $\tau = 0.01, b = 0.9,$ and $K_{dc} = 0.01$ represent, respectively, the sampling time, the motor torque and the back electromotive force. Moreover, $[\nu_1;\nu_2] \in U, \nu_1,\nu_2 \in \{-0.3,-0.25,-0.2,\dots, 0.2,0.25,0.3\}$ are control inputs. We assume that the model is unknown. The main goal is to construct a symbolic model together with an ABF from data by solving $\text{SOP}$ in~\eqref{SOP}. We then employ the data-driven symbolic model as a suitable substitute of original system and synthesize policies maintaining states of unknown DC motor in a safe set $X = [-0.5,0.5]^2$ for infinite time horizons with some guaranteed confidence level.

We fix the structure of our ABF as $\mathcal V(\eta,x,\hat x) = \eta_1(x_1 - \hat x_1)^4 + \eta_2(x_2 - \hat x_2)^4 + \eta_3$. We also fix the threshold $\varepsilon_k = 0.013$, $\forall k\in\{1,\dots,l\}$, and the confidence $\beta = 10^{-2}$, a-priori. Now we need to compute $\mathcal I_{\phi_k}$ which is required for computing the minimum number of data. We construct matrix $P$ based on coefficients of ABF. By considering each coefficient of ABF between $[-0.2 , 0.2]$, we ensure that $\lambda_{\max}(P) \le 0.4$ as discussed in Remark~\ref{Gerschgorin}. We a-priori fix $\delta = 0.05$. We also assume that $\tilde\gamma \in \{0.1,0.2,0.3\}$ with the cardinality $l = 3$. Then according to Lemma~\ref{Lem:1}, we compute  $\mathcal I_{\phi_1} = \mathcal I_{\phi_2} = \mathcal I_{\phi_3} = 1.55$. Since the number of decision variables affects the minimum required number of data in~\eqref{EQ:12}, we fix $\tilde\rho = 0.015$ a-priori, to reduce the number of decision variables to $4$. Now we have all the required ingredients to compute $\mathcal N$. The minimum number of data required for solving $\text{SOP}$ in~\eqref{SOP} is computed as $\mathcal N = 156052$.
We now solve the $\text{SOP}$~\eqref{SOP} with the acquired $\mathcal N$ and the additional conditions on coefficients of the nominated ABF. Coefficients of ABF together with the optimal objective value of $\text{SOP}$ are computed as
\begin{align*}
\mathcal V(\eta,x,\hat x) = 0.2(x_1 - \hat x_1)^4 + 0.01(x_2 - \hat x_2)^4 + 0.2,~~\mu_{\mathcal N}^*=-0.014.
\end{align*}
Since $\mu^*_{\mathcal N} + \max_k\varepsilon_k = -10^{-3} \leq 0$, according to Theorem~\ref{Thm:3}, the constructed $\mathcal V$ from data is an ABF between symbolic model $\hat\Upsilon$ and unknown DC motor $\Upsilon$ with $\gamma = 0.99,\rho = 0.021$ and with a confidence of at least $99\%$. Then according to Theorem~\ref{Thm:1}, the relation $\mathscr{R} \subseteq X \times \hat X$ defined by $\mathscr{R} := \big\{(x,\hat x) \in X \times \hat X \,\big|\, \mathcal V(x,\hat x) \leq  0.021\big\}$
is an $\tilde\epsilon$-approximate alternating bisimulation relation between $\hat\Upsilon$ and $\Upsilon$ with $\tilde\epsilon = 0.37$ and a confidence of at least $99\%$.

\begin{figure}[h]
	\centering 
	\includegraphics[width=0.45\linewidth]{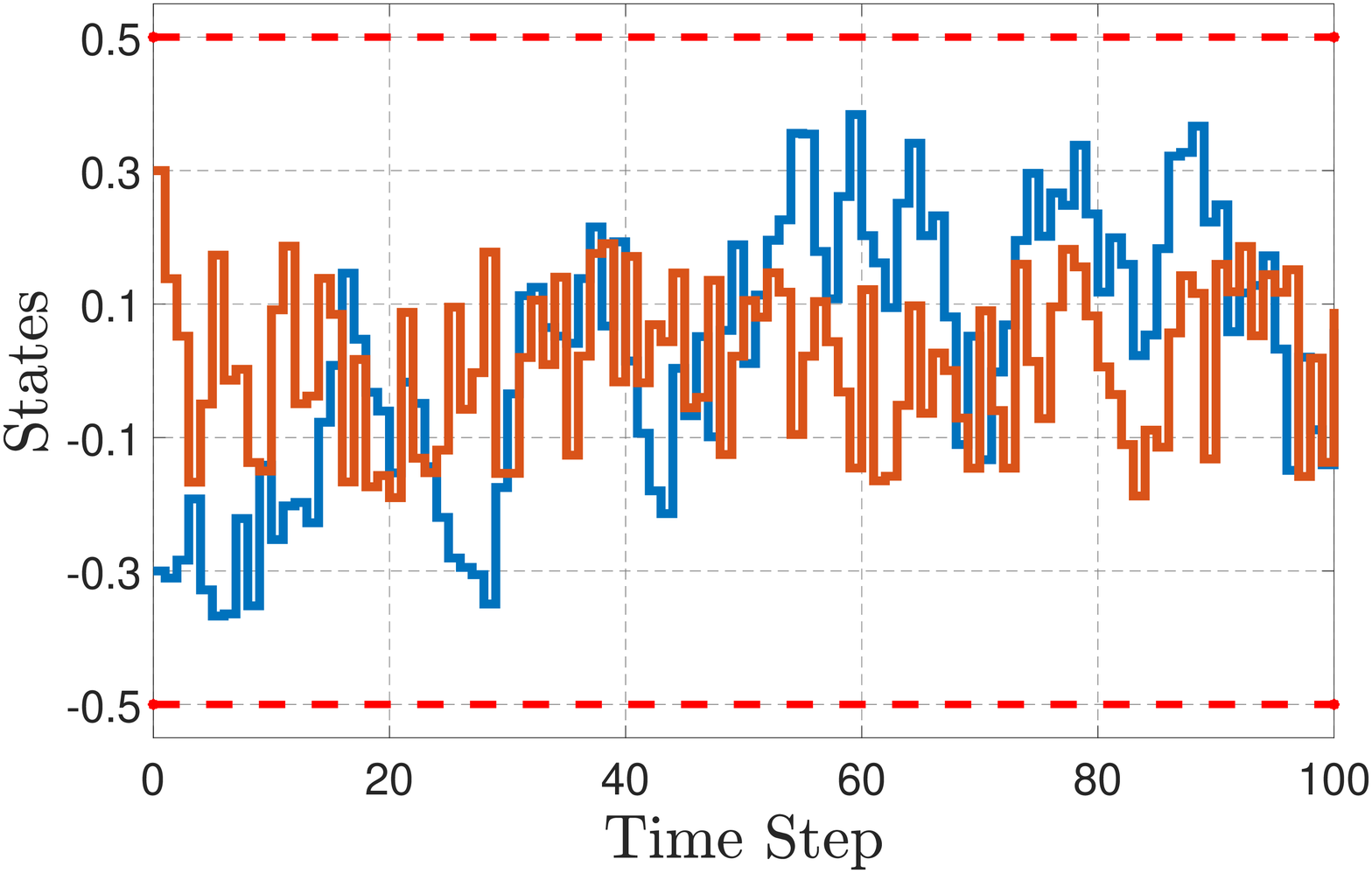}
	\includegraphics[width=0.45\linewidth]{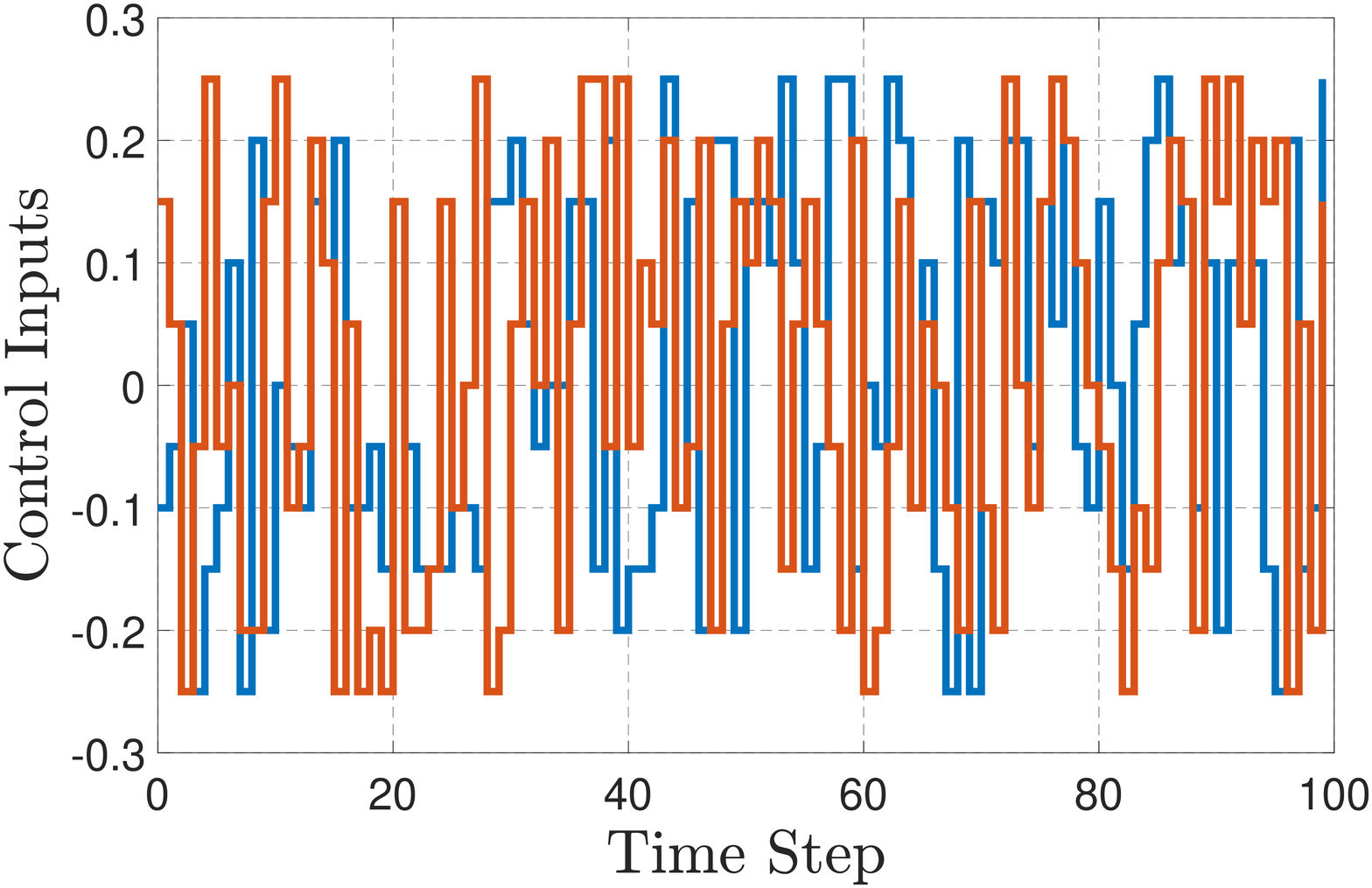}
	\caption{Closed-loop state and input trajectories of unknown DC motor by deploying the synthesized controller from the data-driven symbolic model. Red and blue lines are (state and input) trajectories of the system starting from initial conditions $x_1(0) = 0.3$ and $x_2(0) = -0.3$, respectively. As observed, the synthesized controller keeps the trajectories of unknown DC motor within $[-0.5,0.5]^2$.}
	\label{Simulation1}
\end{figure}

Let us now employ the constructed symbolic model from data and synthesize a controller for $\Upsilon$ via its data-driven symbolic model $\hat \Upsilon$ such that the controller maintains states of the DC motor in the comfort zone $[-0.5,0.5]^2$. We employ the tool \texttt{SCOTS} \cite{rungger2016scots} to synthesize controllers for $\hat \Upsilon$. Closed-loop state and input trajectories of DC motor are illustrated in Fig.~\ref{Simulation1}. As it can be observed, the synthesized controller keeps the trajectories of unknown DC motor within $[-0.5,0.5]^2$. 

{\bf Jet Engine Compressor.} In order to show the applicability of our results to \emph{nonlinear} systems, we apply our approaches to the following nonlinear jet engine compressor~\cite{anta2010sample}:
\begin{align}\notag
x_1(k+1) &= x_1(k) + \tau\big (-x_2(k) - \frac{3}{2}x_1^2(k) - \frac{1}{2}x_1^3(k)\big ),\\\label{jet}
x_2(k+1) &= x_2(k) + \tau\big (x_1(k) - \nu(k)\big ),
\end{align}
where $x_1 = \Gamma - 1, x_2 = \tilde\Gamma - \hat\Gamma - 2$, with $\Gamma,\tilde\Gamma,\hat\Gamma$ being, respectively, the mass flow, the pressure rise, and a constant, $\nu \in U = \{-0.5,-0.45,-0.4,\dots, 0.4,0.45,0.5\}$ is the control input, and $\tau = 0.01$ is the sampling time.
We assume that the model is unknown. 
The main goal is to construct a symbolic model together with an ABF from data. We then employ the data-driven symbolic model and synthesize policies maintaining states of unknown jet engine in a safe set $X = [-0.5,0.5]^2$ for infinite time horizons with a guaranteed confidence level.

Let us fix the structure of our ABF as $\mathcal V(\eta,x,\hat x) = \eta_1(x_1 - \hat x_1)^2 + \eta_2(x_2 - \hat x_2)^2 + \eta_3$. We also fix the threshold $\varepsilon_k = 6\times 10^{-3}$, $\forall k\in\{1,\dots,l\}$, and the confidence $\beta = 10^{-2}$. By considering coefficient of ABF between $[-0.2 , 0.2]$, we ensure that $\lambda_{\max}(P) \le 0.4$. We a-priori fix $\delta = 0.05$ and assume that $\tilde\gamma \in \{0.1,0.2,0.3\}$ with the cardinality $l = 3$. Then according to Lemma~\ref{Lem:1_1}, we compute  $\mathcal I_{\phi_1} = 1.8, \mathcal I_{\phi_2} = 1.9,\mathcal I_{\phi_3} = 2.02$. We also fix $\tilde\rho = 0.01$ a-priori, to reduce the number of decision variables to $4$. Then the minimum number of data required for solving $\text{SOP}$ in~\eqref{SOP} is computed as $\mathcal N=1100794$. We now solve the $\text{SOP}$~\eqref{SOP} with the obtained $\mathcal N$ and the additional conditions on coefficients of the nominated ABF. Coefficients of ABF together with the optimal objective value of $\text{SOP}$ are computed as
\begin{align*}
\mathcal V(\eta,x,\hat x) = 0.0004(x_1 - \hat x_1)^2 + 0.0016(x_2 - \hat x_2)^2 + 0.2,~~\mu_{\mathcal N}^*=-0.007.
\end{align*}
Since $\mu^*_{\mathcal N} + \max_k\varepsilon_k = -10^{-3} \leq 0$, according to Theorem~\ref{Thm:3}, the constructed $\mathcal V$ from data is an ABF between symbolic model $\hat\Upsilon$ and unknown jet engine $\Upsilon$ with $\gamma = 0.99,\rho = 0.012$ and with a confidence of at least $99\%$. Then according to Theorem~\ref{Thm:1}, the relation $\mathscr{R} := \big\{(x,\hat x) \in X \times \hat X \,\big|\, \mathcal V(x,\hat x) \leq 0.012\big\}$
is an $\tilde\epsilon$-approximate alternating bisimulation relation between $\hat\Upsilon$ and $\Upsilon$ with $\tilde\epsilon = 0.35$ and a confidence of at least $99\%$.

\begin{figure}[h!]
	\centering 
	\includegraphics[width=0.45\linewidth]{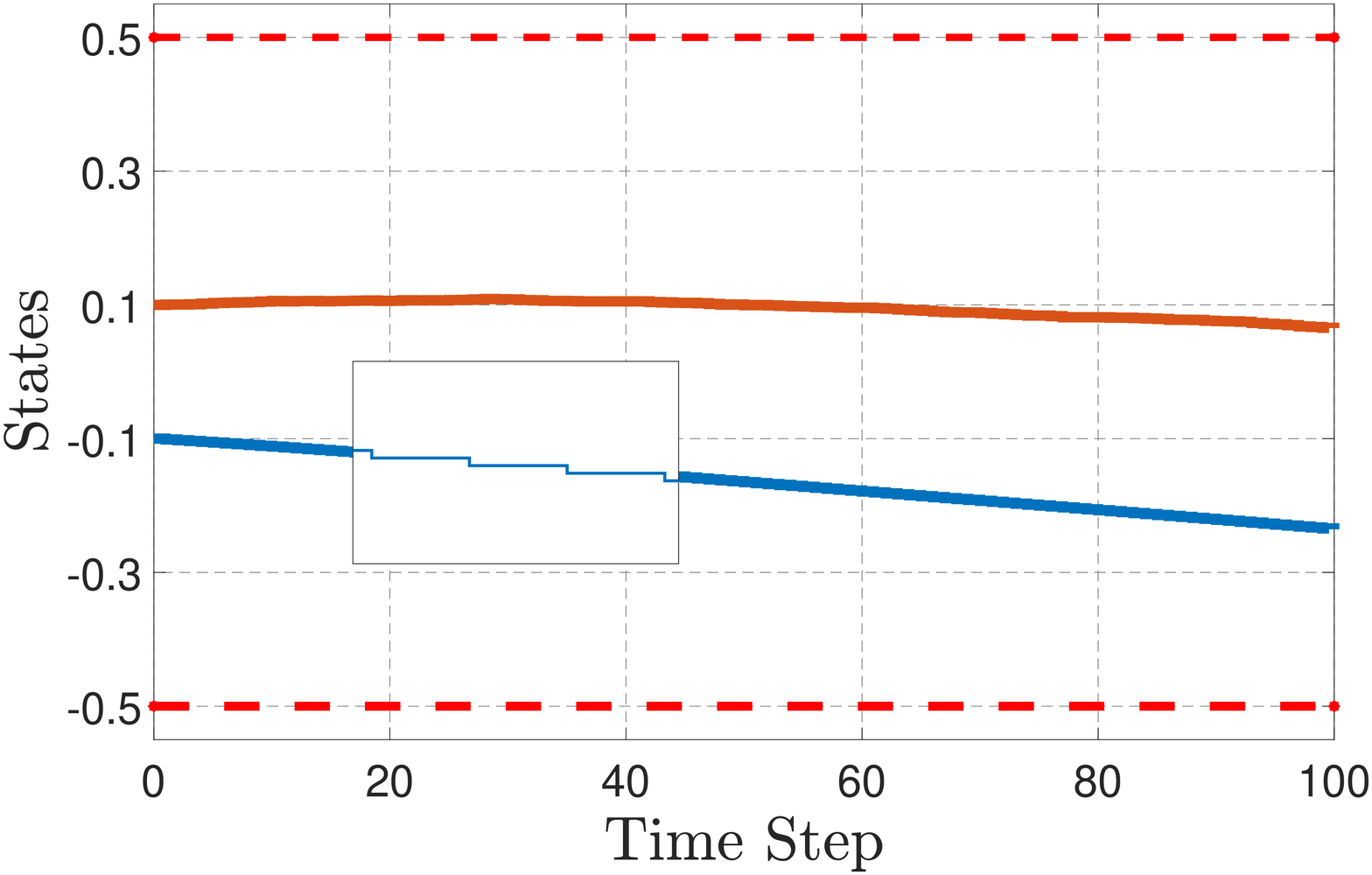}
	\includegraphics[width=0.45\linewidth]{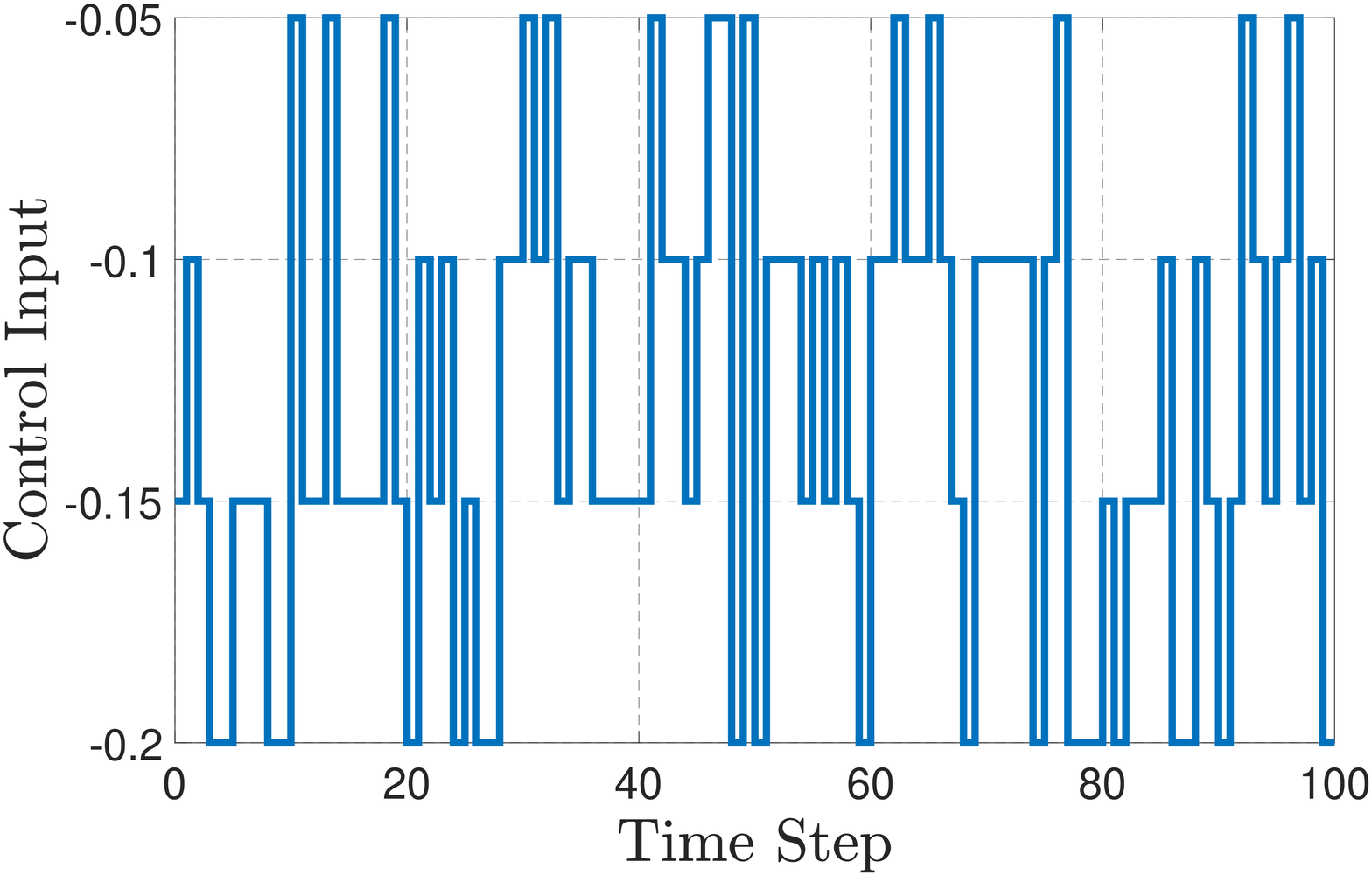}
	\caption{Closed-loop state and input trajectories of unknown jet engine by deploying the synthesized controller from the data-driven symbolic model. Red and blue lines are (state and input) trajectories of the system starting from initial conditions $x_1(0) = 0.1$ and $x_2(0) = -0.1$, respectively. There is no red line in the right figure since the control input only exists in state $x_2$.}
	\label{Simulation}
\end{figure}

We now synthesize a controller for $\Upsilon$ via its data-driven symbolic model $\hat\Upsilon$ such that the controller maintains states of the jet engine in the comfort zone $[-0.5,0.5]^2$. Closed-loop state and input trajectories of the jet engine compressor are illustrated in Fig.~\ref{Simulation}. As observed, the synthesized controller keeps the trajectories of unknown jet engine compressor within $[-0.5,0.5]^2$. It is worth mentioning that in both case studies, we considered basis functions $q_j(x,\hat x)$ to be monomials over $x,\hat x$, and accordingly, the ABF to be polynomial since unknown DC motor and jet engine models are going to be polynomial based on the physics.

\begin{figure}[h!]
	\centering
	\includegraphics[scale=0.27]{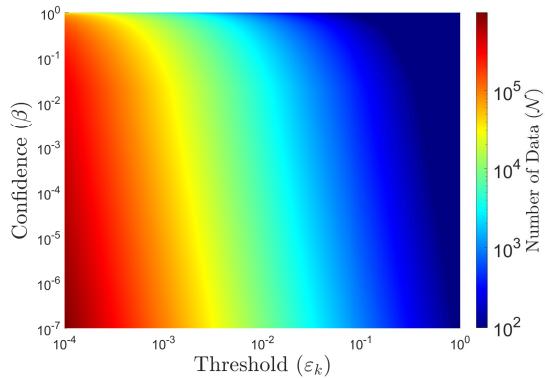}
	\caption{Required number of data, represented by `colour bar', in terms of the threshold $\varepsilon_k$ and the confidence $\beta$. Plot is in the logarithmic scale. The required number of data decreases by increasing either the threshold $\varepsilon_k$ or the confidence $\beta$.}
	\label{Fig3}
\end{figure}

{\bf Computational Complexity Analysis.} In order to provide a more practical analysis on the computational complexity based on number of collected data required for solving the scenario optimization program in~\eqref{SOP}, we plotted in Fig.~\ref{Fig3} the required number of data in terms of the threshold $\varepsilon_k$ and the confidence $\beta$ based on~\eqref{EQ:12} for the jet engine. As it can be observed, the required number of data decreases by increasing either the threshold $\varepsilon_k$ or $\beta$.

\section{Conclusion}
In this work, we proposed a data-driven approach for the construction of symbolic models for unknown deterministic control systems. We utilized
alternating bisimulation functions, as
a relation between each unknown system and its symbolic model, to quantify the mismatch between state behaviors of two systems. In our proposed setting, we cast the required conditions for constructing ABF as a robust optimization program (ROP) and then proposed a scenario optimization program (SOP) associated to the original ROP. We established a probabilistic bridge between optimal values of SOP and ROP, and formally constructed ABF between unknown systems and their symbolic models with a guaranteed confidence level. We demonstrated our approaches over two physical case studies with unknown dynamics including (i) a DC motor and (ii) a \emph{nonlinear} jet engine compressor. Developing a \emph{compositional} approach for the data-driven construction of symbolic models for \emph{large-scale} dynamical systems is under investigation as a future work.

\bibliographystyle{alpha}
\bibliography{biblio}

\section{Appendix}

{\bf Proof of Theorem~\ref{Thm:3}.} We first establish a probabilistic
relation between optimal values of ROP and SOP. Based on~\cite[Theorems 4.1, 4.3]{esfahani2014performance}, the
probabilistic distance between optimal values of
ROP and SOP can be formally lower bounded as\footnote{One can readily verify that $\mu^*_{\mathcal R}$ is always bigger than or equal to $\mu^*_{\mathcal N}$ since $\mu^*_{\mathcal R}$ is computed for infinitely many constraints whereas $\mu^*_{\mathcal N}$ is computed only for finitely many of them.}
\begin{align}\label{EQ:12_1}
\PP^{\mathcal N} \Big\{0\leq\mu^*_{\mathcal R}-\mu^*_{\mathcal N}\leq\max_k\varepsilon_k\Big\}\geq 1-\beta,
\end{align}
provided that $$\mathcal N\geq \mathcal {\bar N}\big(w(\frac{\varepsilon_k}{\mathrm L_{\mathrm {SP}}\mathcal I_{\phi_k}}),\beta\big),$$ where
$w\!: [0,1]\rightarrow [0,1]$ is given by
\begin{align}\notag
w(s) = s^{n}, \quad \forall s \in  [0,1],
\end{align}
and $\mathrm{L}_{\mathrm{SP}}$ is a Slater point as defined in~\cite[equation (5)]{esfahani2014performance}. Since the original ROP in~\eqref{ROP} is a $\min$-$\max$ optimization problem, the Slater constant  $\mathrm{L}_{\mathrm{SP}}$ can be selected as 1~\cite[Remark 3.5]{esfahani2014performance}. We refer the interested reader to ~\cite[equation (5)]{esfahani2014performance} for more details on the formal definition of Slater constant.

\noindent
From~\eqref{EQ:12_1}, one can readily conclude that $\mu^*_{\mathcal N}\leq\mu^*_{\mathcal R}\leq\mu^*_{\mathcal N}+\max_k\varepsilon_k$ with a confidence of $1-\beta$. If $\mu^*_{\mathcal N}+\max_k\varepsilon_k \leq 0$ (as the
main condition of the theorem), then $\mu^*_{\mathcal R} \leq 0$, implying that  conditions~\eqref{EQ:5}-\eqref{EQ:6} are satisfied and the constructed $\mathcal V$ is an ABF between $\hat\Upsilon$ and $\Upsilon$ with a confidence of at least $1-\beta$, which completes the proof. $\hfill\blacksquare$

{\bf Proof of Lemma~\ref{Lem:1}.} We first compute Lipschitz constants of $\phi_1, \phi_2$ with respect to $x$ and then take the maximum between them. For $\phi_2$, we have
\begin{align}\notag
\mathcal I_{{2_k}}=\max\limits_{x\in X, \Vert x\Vert \leq \alpha_1}\Vert\frac{\partial \phi_{2}}{\partial x}\Vert.
\end{align}
Accordingly, 
\begin{align*}
&\mathcal I_{{2_k}} = \max\limits_{x\in X, \Vert x\Vert \leq \alpha_1}\Vert 2((Ax + B\nu) - \mathcal Q(A\hat x+ B\nu))^\top PA - 2\tilde\gamma_k(x-\hat x)^\top P\Vert\\
&\leq \max\limits_{x\in X, \Vert x\Vert \leq \alpha_1}\Vert  2((Ax + B\nu) - \mathcal Q(A\hat x+ B\nu))^\top PA\Vert +\Vert 2\tilde\gamma_k(x-\hat x)^\top P\Vert\\
& \leq  \max\limits_{x\in X, \Vert x\Vert \leq \alpha_1} 2\Vert P \Vert\big(\Vert A \Vert(\Vert A x\Vert + \Vert B  \nu \Vert + \Vert \mathcal Q(A\hat x + B \nu)\Vert)+\tilde\gamma_k(\Vert x \Vert + \Vert \hat x \Vert)\big)\\
& \leq  \max\limits_{x\in X, \Vert x\Vert \leq \alpha_1} 2\Vert P \Vert\big(\Vert A \Vert(\Vert A x\Vert + \Vert B  \nu \Vert + \delta + (\Vert A\hat x + B \nu \Vert))+\tilde\gamma_k(\Vert x \Vert + \Vert \hat x \Vert)\big)\\
& \leq 2\lambda_{\max}(P) (2\mathscr I_1^2 \alpha_1 + 2\mathscr I_1 \mathscr I_2 \alpha_2 + \mathscr I_1\delta + 2\alpha_1\tilde\gamma_k).
\end{align*}
For computing $\mathcal I_{1}$, since $\lambda_{\min}(P)\Vert x-\hat x \Vert^2 \leq (x-\hat x)^\top P(x-\hat x)$, we have $\sigma = \lambda_{\min}(P)$ in~\eqref{EQ:11}. Then one has
\begin{align*} 
\mathcal I_{1} =\max\limits_{x\in X, \Vert x\Vert \leq \alpha_1}\Vert 2\lambda_{\max}(P) (x - \hat x) - 2P(x-\hat x)\Vert\leq 4\alpha_1 (\lambda_{\max}(P) + \lambda_{\max}(P)).
\end{align*}

Then $\mathcal I_{\phi_k} \!=\! \max\! \big\{\mathcal I_{1}, \mathcal I_{{2_k}}\big\} \!=\! \max\!\big\{2\lambda_{\max}(P) (2\mathscr I_1^2 \alpha_1 + 2\mathscr I_1 \mathscr I_2 \alpha_1 + \mathscr I_1\delta + 2\alpha_1\tilde\gamma_k),4\alpha_1 (\lambda_{\max}(P) + \lambda_{\max}(P))\big\}$, which completes the proof.$\hfill\blacksquare$

{\bf Proof of Lemma~\ref{Lem:1_1}.} For $\phi_2$, we have
\begin{align*}
&\mathcal I_{{2_k}} = \max\limits_{x\in X, \Vert x\Vert \leq \alpha_1}\Vert  2(f(x,\nu) - \mathcal Q(f(\hat x,\nu))^\top P\partial_{x}f(x,\nu) - 2\tilde\gamma_k(x-\hat x)^\top P\Vert\\
& \leq  \max\limits_{x\in X, \Vert x\Vert \leq \alpha_1} 2\Vert P \Vert\big(\Vert\partial_{x}f(x,\nu)\Vert(\Vert f(x,\nu)\Vert + \delta + \Vert f(\hat x,\nu)\Vert) + \tilde\gamma_k(\Vert x \Vert + \Vert \hat x \Vert)\big)\\
& \leq 2\lambda_{\max}(P) (2\mathscr I_f\mathscr I_x + \mathscr I_f\delta + 2\alpha_1\tilde\gamma_k).
\end{align*}
For $\phi_1$:
\begin{align*} 
\mathcal I_{1} =\max\limits_{x\in X, \Vert x\Vert \leq \alpha_1}\Vert 2\lambda_{\max}(P) (x - \hat x) - 2P(x-\hat x)\Vert \leq 4\alpha_1 (\lambda_{\max}(P) + \lambda_{\max}(P)).
\end{align*}
Then $\mathcal I_{\phi_k} = \max \big\{\mathcal I_{1}, \mathcal I_{{2_k}}\big\} = \max\big\{2\lambda_{\max}(P) (2\mathscr I_f\mathscr I_x + \mathscr I_f\delta + 2\alpha_1\tilde\gamma_k),4\alpha_1 (\lambda_{\max}(P) + \lambda_{\max}(P))\big\}$, which completes the proof.$\hfill\blacksquare$

\end{document}